\documentclass{article}
\usepackage[margin=1in]{geometry}

\usepackage{lineno}
\usepackage{svg}
\usepackage{caption}
\usepackage{subcaption}
\usepackage[version=4]{mhchem}
\usepackage{derivative}
\usepackage{color, soul}
\usepackage{textcomp}
\usepackage{cleveref}
\usepackage{nicefrac}
\usepackage{changepage}
\usepackage{authblk}
 
\begin{document} 

\captionsetup[subfigure]{justification=justified,singlelinecheck=false}
\newcommand{\varsizetwo}{20pt}
\newcommand{\varsizeone}{1.45in}

\title{Low Loss Aluminum Nitride Waveguide Fabrication: Propagation Loss Reduction Through ALD and RTA}

\author[a,b]{Nikolay Videnov\footnote{nikolay.videnov@uwaterloo.ca}}
\author[a,c]{Matthew L. Day}
\author[a,b,c]{Michal Bajcsy}

\renewcommand{\Affilfont}{\small \itshape}

\affil[a]{Institute for Quantum Computing, 200 University Ave W, Waterloo, ON N2L 3G1}
\affil[b]{Department of Electrical and Computer Engineering, 200 University Ave W, Waterloo, ON N2L 3G1}
\affil[c]{Department of Physics and Astronomy, 200 University Ave W, Waterloo, ON N2L 3G1}

\maketitle

\renewenvironment{abstract}
 {\small 
  \list{}{%
    \setlength{\leftmargin}{1in}% <---------- CHANGE HERE
    \setlength{\rightmargin}{\leftmargin}%
  }%
  \item\relax}
 {\endlist}

\begin{abstract} 
\textbf{
Aluminum nitride (AlN) has emerged as a leading platform for integrated photonics in the visible and ultraviolet spectral ranges, particularly for quantum information applications involving trapped atoms. However, achieving low propagation loss in tightly confining single-mode AlN waveguides remains a challenge, especially at sub-micron wavelengths where scattering losses scale unfavorably. In this work, we present a reproducible and detailed fabrication process for low-loss AlN waveguides on sapphire, achieving a record loss of 2~dB/cm at 852~nm. Our results are enabled by systematic process optimization including high-resolution electron beam lithography with shape-based proximity effect correction, atomic layer deposition (ALD) of \ce{Al2O3} for waveguide surface passivation, and post-fabrication rapid thermal annealing (RTA). We provide a study of the contributions of each technique to propagation loss reduction and support our findings with quantitative loss measurements and comparison with an exhaustive literature review. This work represents the first detailed report of ALD passivation and post-cladding RTA applied to AlN waveguides.
}
\end{abstract}

\section{Introduction}\label{sec:intro}
Silicon photonic integrated circuits (PIC) have become a mature technology \cite{Rickman2014}. Consequently, at telecommunication wavelengths there is broad availability for turn-key optical components. However, silicon becomes opaque at 1.1~$\mu$m barring many areas of research from leveraging these advancements in PIC technology; fields as diverse as trapped atom quantum technologies, underwater LIDAR, and biological imaging all stand to benefit immensely from a UV-VIS PIC industry \cite{Park2024}. This has spurred recent interest in large band gap alternatives to silicon, as these materials boast extended transparency windows that accommodate shorter wavelengths.

Presently, there are two popular materials with a large band gap: silicon nitride (\ce{SiN_x}) and lithium niobate (\ce{LiNbO3}), which both have a 4~eV (310~nm) band gap and a maturing fabrication ecosystem with a large variety of devices having been already demonstrated \cite{Xiang2022}\cite{Zhu2021}. For deep UV ($<$300~nm) or high power applications, \ce{SiN_x} and \ce{LiNbO3} are still insufficient due to direct band gap absorption or two-photon absorption. To extend the operation of UV-VIS PIC even further, two alternative materials are being explored: aluminum oxide (\ce{Al2O3}) \cite{McKay2023} and aluminum nitride (\ce{AlN}) \cite{Liu2023}. Of the two, \ce{AlN}, is the only high index (2.2), large band gap (6~eV, 200~nm),  electro-optically, and piezo-electrically active material. Consequently, AlN has been used for non-linear optics \cite{Liu2021}\cite{Weng2021}, electro-optic modulation\cite{Xiong2012nanoLetters}, acousto-optic modulation\cite{Tadesse2014}, hybrid ECDLs \cite{Videnov2024}, and much more \cite{Liu2023}.

Despite these exciting results in AlN, the performance of many of these devices is limited by the propagation loss. This is a universal problem in PIC, with propagation loss being a key metric for any PIC platform. To that end, though the material properties of the platform are critical for motivating the use and study of the material, having the ability to fabricate low loss waveguides is what enables the fabrication of exciting new devices and ultimately pushes forward the material's popularity. 

In this paper we describe a recipe for low loss aluminum nitride waveguides in the full detail needed to replicate our results. We begin with a literature review of AlN waveguides in Section~\ref{sec:stateoftheart}. Next, we detail the fabrication steps for our waveguides in Section~\ref{sec:recipe}, with more detailed discussion of three key steps: electron beam lithography in Section~\ref{sec:jeol-beamer}, atomic layer deposition in Section~\ref{sec:ALD}, and rapid thermal annealing in Section~\ref{sec:RTA}; in each section the focus is on the effect on propagation loss. We hope that these results will aid in reducing waveguide losses within the AlN PIC community.

\section{Aluminum Nitride Waveguides}\label{sec:stateoftheart}

\nocite{Lu2018,Liu2018,Jung2016, Alvarado2014, Pernice2012, Zhu2019, Liu2017, Sun2019, Liu2018a, Spettel2024, Xiong2012, Jang2023, Wu2020, Amrar2025, Lin2014, Dong2019}
\begin{figure}[thb]
    \centering
    \includegraphics[width=0.8\textwidth]{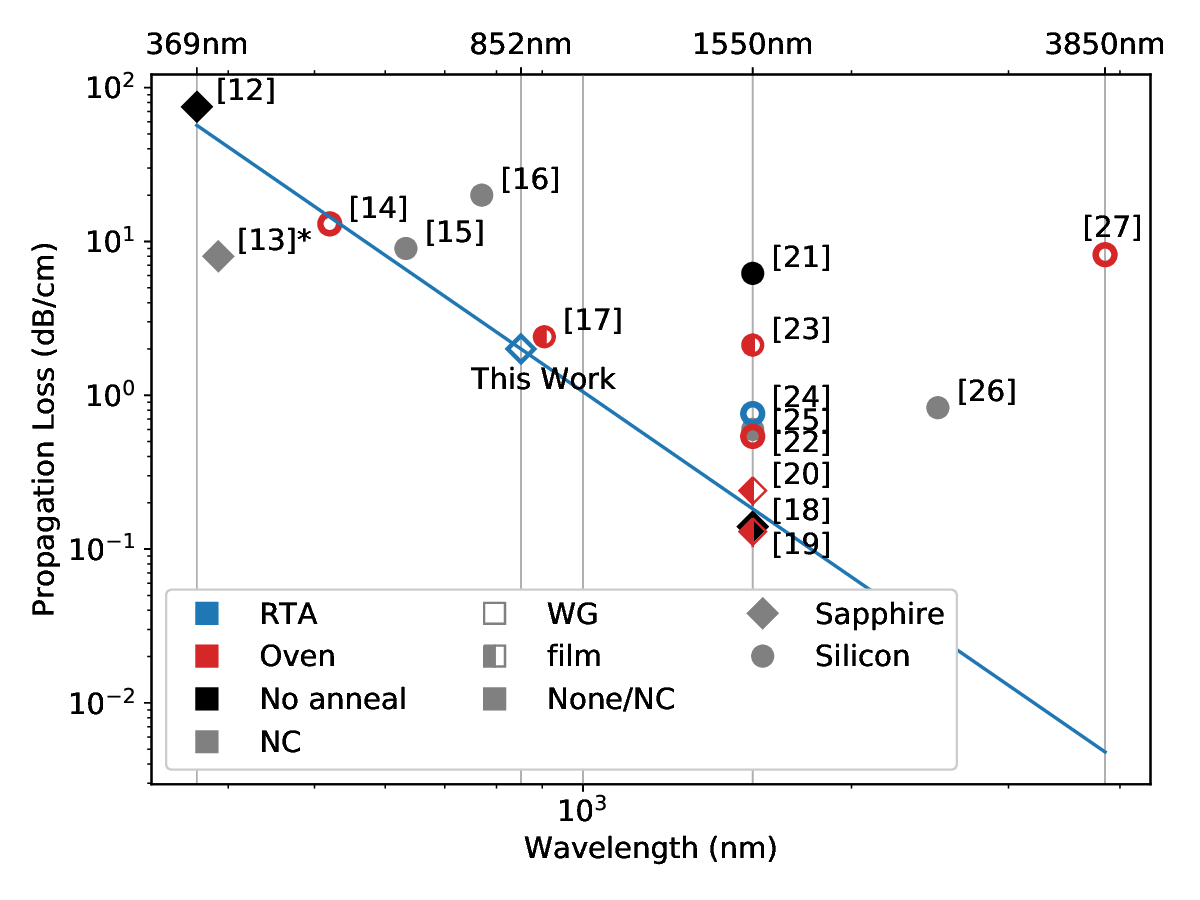}
    \caption{Propagation loss for single mode tightly confining aluminum nitride waveguides across the literature. The solid line is a $1/\lambda^4$ equivalent loss curve which assumes a Rayleigh scattering dominant loss mechanism. The type of annealing, if any, is indicated by the marker color (NC indicates not commented in the reference). The fill of the marker (full, half, or none) shows whether the annealing was performed on the AlN film or on the waveguides after cladding with full indicating no annealing or no information. Lastly, the shape of the marker indicates if the substrate is sapphire or silicon. \cite{Liu2018}* results use a few mode waveguide. \cite{Spettel2024} uses AlN doped with Sc.}
    \label{fig:lossReview}
\end{figure}

We report a propagation loss of 2.0$\pm$0.3~dB/cm at 852~nm, a loss figure outperforming the previously reported best losses around 852~nm, and consistent with the best reported loss at any wavelength. In our structures, sidewall roughness introduced during fabrication is the primary source of loss; material absorption is negligible because 852~nm is far from the AlN band gap and we use high quality commercial thin films. Fig.~\ref{fig:lossReview} summarizes reported loss values in tightly confining, single-mode AlN waveguides from the existing literature. In such waveguides, the optical mode necessarily has significant overlap with the sidewalls, making this a good metric for waveguide fabrication quality. In contrast, multi-mode waveguides tend to confine the fundamental mode well within the core, limiting sidewall interaction. Similarly, in weakly confining waveguides the optical mode resides mostly in the cladding, diminishing the influence of sidewall roughness. While these strategies are effective for minimizing loss, they do not necessarily reflect the intrinsic quality of the waveguide fabrication process.

Waveguide loss has three primary components of which one is usually dominant: material losses, geometric losses, and scattering losses. The first two are addressed by using commercial high quality films and suitable waveguide geometry\footnote{In this case, as determined through Lumerical MODE simulation. A subset of geometric losses is radiation losses in bends, however, here we are well above the critical bend radius eliminating this consideration.}. The last consideration ---scattering losses--- is typically the most important. In the absence of atomically flat sidewalls, the nano-scale roughness allows for the guided light to undergo scattering. The fabrication challenge is further exacerbated when using shorter wavelengths because the scattering losses have a unfavorable wavelength dependence.

Scattering losses are commonly modeled as Rayleigh scattering from sidewall roughness, which has a $1/\lambda^4$ wavelength dependence \cite{Patel2023}. In Fig.~\ref{fig:lossReview} we observe that the best performing recipes across the spectrum conform with a $1/\lambda^4$ scaling, considering only tightly confining single mode waveguides, which supports the claim that Rayleigh scattering is the dominant loss mechanism. Previous studies, using varying wavelengths in a series of devices, have not always found the same scaling, though still ascribing losses to Rayleigh scattering. Corato-Zanarella \emph{et al.} performed an excellent study in \ce{SiN_x} waveguides fabricating material absorption limited waveguides operating at wavelengths from 461~nm to 780~nm, where they found a $1/\lambda^{2.1}$ loss scaling in their devices when separating out material and scattering losses \cite{Corato-Zanarella2024}. Similarly, Liu \emph{et al.} find a $1/\lambda^3$ scaling in their AlN waveguides between 390~nm and 455~nm, though their waveguides were not single mode for the whole wavelength range \cite{Liu2018}.

From this figure, we can also observe that the best propagation losses are generally, though not exclusively, achieved with AlN grown on sapphire. Sapphire is used as a substrate because AlN and sapphire have compatible crystal lattices, minimizing lattice miss-match when growing the film. Furthermore, sapphire has a large band gap, maintaining the AlN transparency window. However, with a sapphire substrate there are substantial challenges not typically discussed. Foremost, the sapphire crystal does not readily cleave into optical facets. In the literature, there are some reports of optical quality facets from cleaving \cite{Liu2018}, but we were not able to replicate this. Gow \emph{et al.} may show the way forward here, with optical quality facet preparation through dicing \cite{Gow2024}. Facet preparation through etching is also frequently employed in similar situations, however, single crystal sapphire is exceptionally durable, which presents a significant challenge for the deep etching needed in facet preparation. A popular alternative to edge couplers is grating couplers, however, they are challenging to fabricate without a buried oxide (BOX) layer. Typically a grating coupler is designed to have destructive interference with the reflection from the BOX-substrate interface. The AlN on sapphire material stack has no BOX-substrate reflection which results in poor efficiency for grating couplers. Efficient bottom-side couplers have been demonstrated in AlN on sapphire to address this issue \cite{Zhou2025}. Lastly, the sapphire layer is not readily etched by any wet etchant, which limits the possibility of suspended devices. 

Every AlN waveguide plotted in Fig.~\ref{fig:lossReview} which has a Si substrate has a BOX layer to separate the small band gap Si from the AlN. With Si as the substrate, high quality facets can be easily produced through cleaving, grating couplers can be fabricated in the usual way, and suspended devices are easily fabricated through HF undercutting. The only drawback is excess loss due to the inferior AlN film quality. In Fig.~\ref{fig:lossReview}, we observe that though the best losses use sapphire as the substrate, there are a handful of results using a Si substrate along the best performance curve, which may even become record loss if they are not applying a RTA and ALD step. It is outside the scope of this paper but worth mentioning that we have not observed any benefit of using UV grade AlN templates over the B grade templates, indicating again that material losses are secondary. The difference between these two grades from Kyma Technologies is whether the AlN is single crystal or polycrystalline in XY while both grades are single crystal along the c-axis. The natural conclusion is that the crystalinity of the AlN is not a major concern at this junction.

In conclusion, we find Rayleigh scattering to be the dominant loss mechanism across the literature for best performing waveguides. Though sapphire substrates are over-represented at the lowest loss numbers, good results can also be achieved with a Si substrate which offers many ancillary benefits. 

\section{Device Fabrication}\label{sec:recipe}
\begin{figure}[thb]
    \centering
    \includegraphics[width=0.8\textwidth]{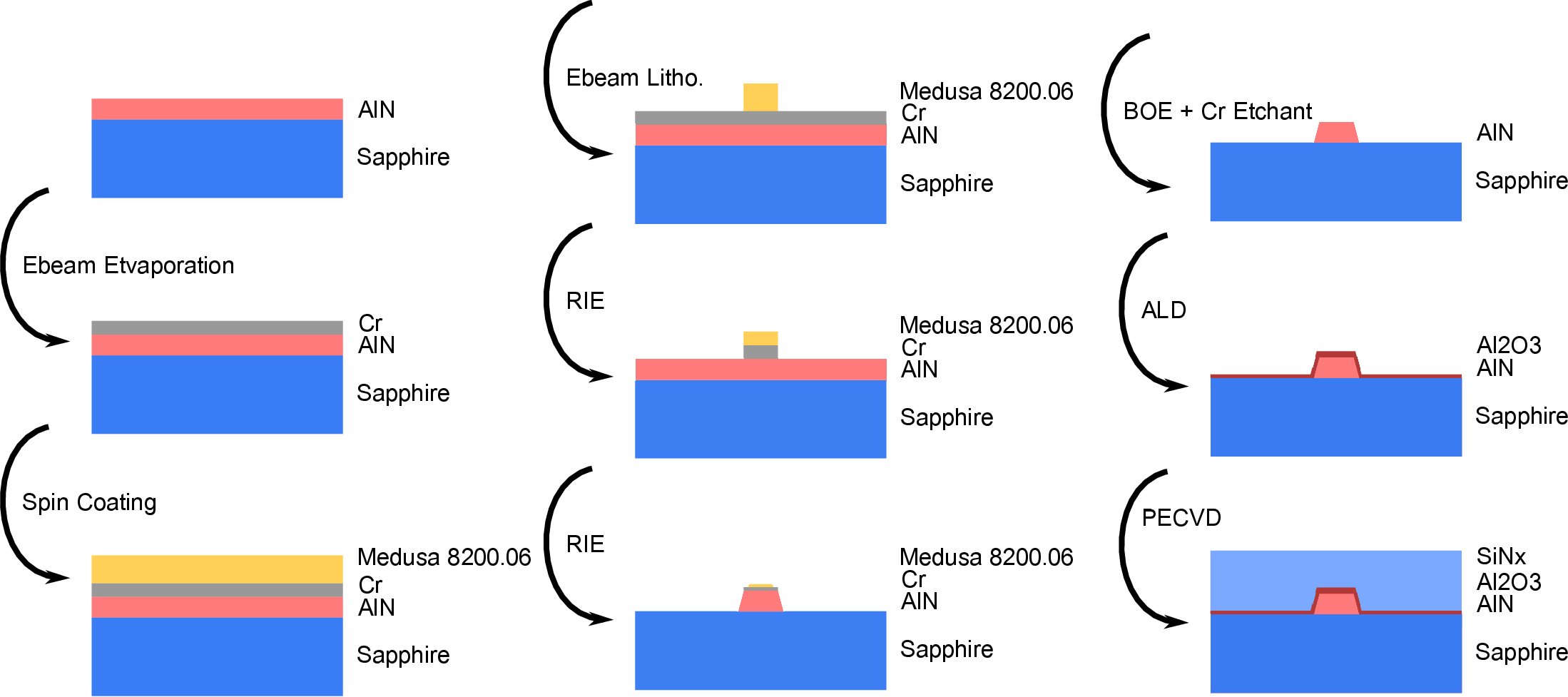}
    \caption{Nano fabrication flow diagram. Subsequent to the final step pictured the chips are diced and polished for edge coupling. The final step is a RTA treatment.}
    \label{fig:fabFlowDiagram}
\end{figure}

B grade AlN on sapphire templates were purchased from Kyma Technologies corresponding to AlN grown through physical vapor deposition with nano-columns. From the purchased wafers 1~cm~x~1~cm pieces were diced using a DISCO DAD3240 dicing saw with a 2.050-8A-54RU7-3 diamond blade from Thermocarbon. To protect the AlN surface during dicing a protective film of AZ P4620 thick photo-resist was spun onto it. The pieces were ultrasonically cleaned after dicing, successively using acetone and IPA. Cleaning was finished with surface ashing using a Yield Engineering Systems CV200RFS (RF power of 1000~W, 400~mTorr, 25~sccm \ce{O2} flow, at 180~\textdegree C for 120~s ).\footnote{No difference was found including or omitting a Piranah clean after ashing, the latter was favored.}

50~nm of Cr metal hard mask which doubled as a discharge layer was deposited onto the AlN using an Angstrom Amod e-beam sputterer with a 3~\AA/s deposition rate. The waveguides were patterned through electron beam lithography using a JBX-6300FS electron beam lithography system to write the patterns in Medusa 8200.06 negative tone resist. The resist was spun on by first exposing the wafer piece to 10~min of UV/Ozone cleaning with a Novascan UV/ozone cleaner followed by a 5~min dehydration bake at 185~\textdegree C. 30~$\mu$L/cm$^2$ of methoxypropanol (PGME) was spun onto the wafer piece and allowed to flash off. On completion of these adhesion promoting preparatory steps, 30~$\mu$L/cm$^2$ of medusa was spun at 5000~rpm, for a typical film thickness of 95~nm, and soft baked at 155~\textdegree C for 10~min\footnote{Manufacturer recipe calls for 150~\textdegree C, this deviation is from a systematic error in hotplate temperature.}. File preparation and exposure details are discussed in Section~\ref{sec:jeol-beamer}. A post exposure bake was performed at 170~\textdegree C for 10~min. Subsequently, the pattern was developed in room temperature AZ~300~MIF developer for 90~s with mild agitation and DI water stop.

The pattern was transferred into the Cr hardmask, then into the AlN, with a two step reactive ion etching (RIE) as described by Table~\ref{tab:etchRecipe}, and performed in an Oxford Instruments ICP380. For an excellent review on etching AlN see the review paper by Pinto \emph{et al.} \cite{Pinto2022}. A short breakthrough Cr etch was found to help resolving small gaps; likely by removing resist residue remaining after development. For the main Cr etch and AlN etch an optical endpoint detection was used based on the intensity of a laser reflected off the wafer piece surface.

Post etch, any remaining electron beam resist residue was removed by a 5~min dip in 10:1 buffered oxide enchant, and the remaining Cr hard-mask was stripped by a 5~min dip in chromium enchant 1020. Atomic layer deposition (ALD) passivation and plasma enhanced chemical vapor deposition (PECVD) cladding were both done in an Oxford System 100 PECVD / FlexAL ALD cluster. First 10~min of UV/Ozone cleaning was performed, then typically 10 cycles of ALD \ce{Al2O3} with TMA precursor were deposited. Subsequently, without breaking vacuum 1.5~$\mu$m of \ce{NH4} free low index \ce{SiN_x} was deposited with PECVD. \ce{SiN_x} was chosen as a cladding material to index match the sapphire substrate.

Optical quality facets were prepared by dicing the wafer pieces 100~$\mu$m offset from the waveguide edge couplers then polishing the resulting facet using an Allied High Tech Products MultiPrep Polishing System with successively finer diamond polishing disks from 30~$\mu$m to 0.1~$\mu$m grit. During polishing the top surface was protected with thick photo-resist. Rapid thermal annealing was performed in an AccuThermo AW 610 rapid thermal annealer, repeated RTA cycles included repeated cleaning of the chips to ensure no contaminate was brought into the RTA chamber.

The completed devices were mounted to a support block using Crystalbond and tested in a waveguide testing station by edge coupling with a fibre array. For more details on the testing station see previous publications by Videnov and Bajcsy \cite{Videnov2025}. Propagation loss was determined by measuring the intensity transmission through a series of loss spirals with known and increasing length. Laser intensity drift was controlled for by monitoring the laser power prior to the device under test with a fibre splitter. 

\begin{table}
    \caption{ICP-RIE Etch Parameters.}
    \label{tab:etchRecipe}
    \begin{adjustwidth}{-.4in}{-.5in}
    \centering
    \begin{tabular}{l|llll|llll|ll}
        \hline\hline
        Step$^a$  & \multicolumn{4}{l|}{Process Gasses (sccm/100)} & Pressure & \multicolumn{2}{l}{Power (W)} & Temp. & DC Bias& Duration$^{b}$\\
        & \ce{BCl3} & \ce{Cl2}& \ce{O2}& \ce{Ar}& (mTorr)& ICP& HF& (\textdegree C) & (V) & (s) \\
        \hline
        Cr Breakthrough & 0 & 47 & 3 & 0 & 3 & 800 & 100 & 50 & 228.3 & 5 \\
        Cr Main Etch & 0 & 47 & 3 & 0 & 6 & 1000 & 20 & 50 & 75 & 110 \\
        AlN Etch & 6 & 36 & 0 & 5 & 3 & 550 & 70 & 20 & 213.4 & 240 \\
        \hline\hline
    \end{tabular}
    \end{adjustwidth}
    $^{a}${Steps are performed sequentially from top to bottom without breaking vacuum.}\\
    $^{b}${The DC bias and duration of the Cr main etch and AlN etch are typical measured parameters. The etch duration of the Cr breakthrough is fixed at 5~s.}
\end{table}

\section{Electron Beam Lithography}\label{sec:jeol-beamer}
\begin{figure}[tbh]
\centering
\includegraphics[width=\textwidth]{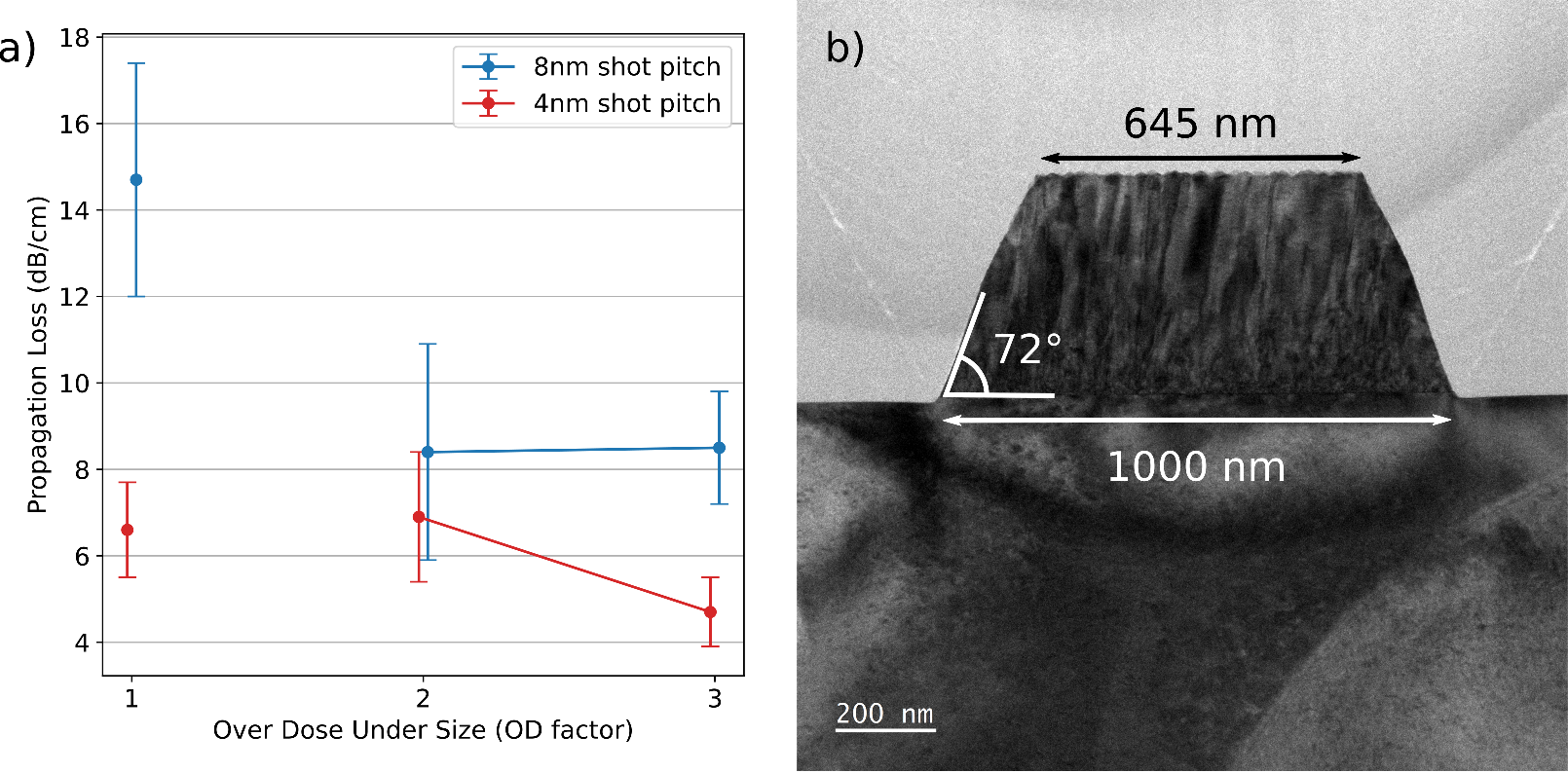}
    \caption{a) EBL file preparation optimization. An over dose (OD) factor of 1 is accepted to mean generic PEC without shape correction, indicated by a discontinuity. b) A TEM cross section of a fabricated waveguide with key as-fabricated dimensions annotated.}
    \label{fig:odus}
\end{figure}

The nominal waveguide dimensions are 750~nm~x~400~nm. These nominal dimensions produce single mode tightly confining waveguides for 852~nm light, and are just under what can reliably be fabricated through direct write UV lithography within the QNFCF facility; consequently the waveguides are patterned by electron beam lithography which easily meets and exceeds the resolution requirements. However, resolution has little relevance for waveguide loss, rather, the sidewall roughness of the mask is more relevant. The ideal etch recipe would reproduce the mask in the wave-guiding material with 1-to-1 accuracy, including any mask roughness. In this section we investigate the effect the mask, patterned through e-beam lithography, has on the waveguide propagation loss.

The electron beam resist is patterned with a JBX-6300FS electron beam lithography (EBL) system, which uses a vector scanning method and provides great flexibility in how patterns can be exposed. The exposure is done point by point by the JEOL on a machine grid. Points on the machine grid can be addressed at a finite rate determined by the control electronics, this is the clock rate which has a maximum of 50~MHz. For many resist doses a dwell time equivalent to 50~MHz will overdose the resist if exposure is done on the machine grid. Hence, a shot pitch is specified which is an even multiple of the machine grid. In this work, the JEOL was used in 4th lens mode with 2~nA beam current and 60~$\mu$m aperture corresponding to a 1~nm machine grid and a minimum 4~nm shot pitch for this recipe. There is also substantial freedom in choosing how the pattern is broken into machine instructions for exposure. All waveguides in this paper were exposed using a 4x multi-pass, and with follow geometry fracturing, to reduce the effect of stitching errors.

Proximity effect correction (PEC) was done in two ways. In both cases, a simulation of the material stack was performed in GenISys TRACER to calculate a point spread function (PSF) for the electron beam which describes how the proximity effect will affect the pattern. The ``generic'' PEC breaks the pattern into trapezoids with a calculated dose correction to account for the finite beam size. If TRACER is allowed to modify the shape of the pattern in addition to applying a dose correction, this is called shape-PEC; an approach which can more accurately correct for the finite beam size, and can be used in an over-dose under-size (ODUS) strategy which improves process latitude \cite{Klimpel2011}. The largely over-dosed resist is less sensitive to variations in post exposure bake, temperature, and development time. In this paper, we pattern waveguides with generic and shape PEC. Both PEC types are patterned with 4~nm and 8~nm shot pitch, and prepared in GenISys BEAMER. For the EBL optimization we patterned each type of PEC and shot pitch sequentially on the same chip, in this way the same recipe is applied to each pattern and the effect of EBL preparation can be isolated.

Across all ODUS settings and both PEC approaches a finer shot pitch results in less propagation loss. We explain this through the finer shot pitch producing smaller amplitude roughness after development. The improvement is between 7~dB/cm and 1~dB/cm, as shown in Fig.~\ref{fig:odus}. The benefit of ODUS compared to generic PEC is less clear; in the 8~nm shot pitch case, shape PEC outperformed the generic PEC by a significant margin. However, for 4~nm shot pitch the benefit appears marginal. Taking all data points together shows a weak trend for higher ODUS factor being beneficial. We do not investigate the process latitude improvements promised by ODUS. A final consideration is that ODUS applies a larger dose to a smaller area, which may allow the use of a finer shot pitch. For the best performing waveguide we used shape PEC with an OD factor of 2.5, and a 4~nm shot pitch.

\section{Atomic Layer Deposition}\label{sec:ALD}

\begin{figure}[tbh]
    \centering
    \includegraphics[width=0.5\linewidth]{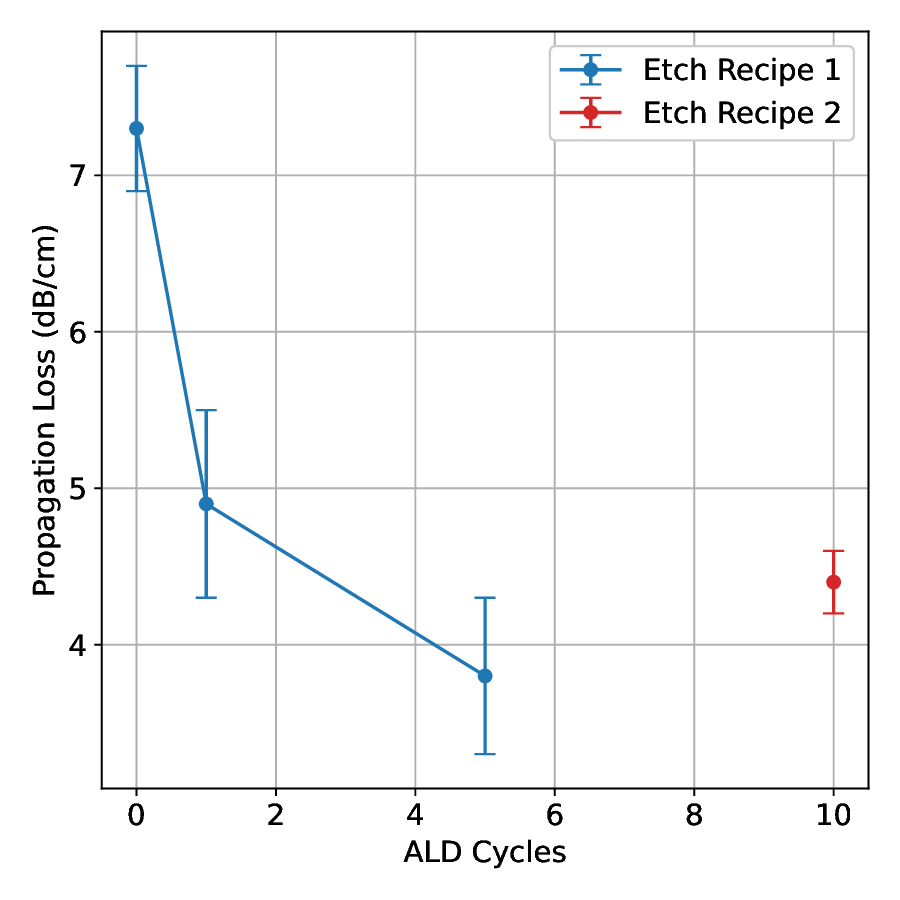}
    \caption{Propagation loss as a function of \ce{Al2O3} ALD cycle count, deposition rate is roughly 1~\AA/cycle. In blue, each sample was prepared the same way aside from ALD count. In red is the best performing chip which had both 10 ALD cycles of \ce{Al2O3}, and an optimized etch recipe. The final jump down to 2.0~dB/cm was only realized after RTA.}
    \label{fig:ALDBenefit}
\end{figure}

Atomic layer deposition (ALD) is a film growth technique which leverages self-limiting chemical processes to deposit single atomic layers of desired materials. By cyclically switching process gasses film thickness can be increased with mono-layer increments \cite{Malygin2015}. Previously, Alasaarela \emph{et al.} have investigated propagation loss improvement through ALD deposition of \ce{TiO2} in air-clad Si waveguides \cite{Alasaarela2011}. They argue that propagation loss is improved due to surface roughness improvement from the conformally coating nature of the ALD deposition process; after many layers of ALD the high frequency roughness is smoothed out.

We found that even a single layer of ALD \ce{Al2O3} significantly improved the propagation loss. As shown in Fig.~\ref{fig:ALDBenefit}, our standard recipe at the time yielded 7.3~dB/cm. With the same etch recipe, waveguide geometry, and testing conditions a single ALD layer reduced the loss to 4.9~dB/cm. Using 5 cycles further reduced the loss to 3.8~dB/cm. Subsequent to these tests, 10 ALD cycles became the standard recipe on the assumption that the loss improvement was asymptotic, 10 cycles corresponds to roughly 1~nm of \ce{Al2O3}. Lumerical MODE simulations show that the effective and group index are affected at the $10^{-3}$ level by 1~nm of \ce{Al2O3}. The negligible index change is confirmed through measurement of the effective refractive index in mirco-ring resonators. For resonators without ALD an effective index of 2.125$\pm$0.004 was measured, while a ring resonator with 5~cycles of ALD was found to have a refractive index of 2.11$\pm$0.1. This difference is on the same order of magnitude as the index variation across the parent wafer from which pieces are diced, as measured by ellipsometry. The expectation is that ALD passivation may reduce the index a small amount. The best performing chip had 10 ALD cycles, however it is important to note that in addition to increasing the ALD count, the etch recipe was significantly different, and an RTA cycle was used to push the loss into the lowest bounds.

The explanation for the observed improvement put forth by Alasaarela \emph{et al.}, a reduction in roughness, may not be valid in this case. The \ce{Al2O3} cladding is closer to matching the \ce{SiN_x} cladding index than it is to the AlN waveguide index. Consequently, the propagating light should still interact with the sidewall roughness at the \ce{Al2O3}-AlN interface. Instead we propose that the mechanism for loss reduction is primarily chemical. After etching the AlN surface may have undesirable dangling functional groups which become defects when cladding with \ce{SiN_x}. The ALD step provides a surface treatment allowing the cladding to be grown with fewer defects at the cladding-waveguide interface. This claim is supported by the observation that as few as 1 ALD cycle yields significant loss improvement, and the rapidly diminishing returns on additional ALD cycles.

These results show that even a single ALD cycle can significantly reduce propagation loss, likely by passivating surface defects. Since ALD and PECVD are often available in the same tool, incorporating ALD passivation is a straightforward and low-cost method of improving waveguide performance---and well worth considering in future fabrication flows.

\section{Rapid Thermal Annealing}\label{sec:RTA}

\begin{table}[h]
\caption{Relevant parameters from previous studies on thermal annealing of AlN for improving waveguide propagation loss$^a$}
\label{tab:rta}
\centering
\begin{tabular}{c|c|c|c|c|cc}
\hline\hline
Year  & Material         & $\lambda_{\text{test}}$     & Fractional Loss& Type & \multicolumn{2}{c}{Optimal}  \\ 
 & &  (nm) &   (L/L$_0$) & & (\textdegree C) & (\textdegree C /s)\\ \hline
2019\cite{Dong2019}      & Si/\ce{SiO2}/AlN/\ce{SiO2} & 3660-3900 & 0.47      & Oven & 400  &  N/A     \\
2020\cite{Wu2020}      & Si/\ce{SiO2}/AlN/\ce{SiO2} & 1550      & 0.58$^b$    & RTA  &  800   &  4.7    \\
This Work & Sapphire/AlN/\ce{SiN_x}     & 852       &  0.45$\pm$0.07   & RTA  &  400 & 6.3 \\ \hline\hline
\end{tabular}
\newline
\begin{flushleft}
$^a${Here we compare the best performing samples.}\\
$^b${In this work the loss is not reported for zero RTA cycles, therefor we compare the fractional improvement between 1 RTA cycle and their optimal loss after 8 RTA cycles which likely underestimates their best fractional improvement.}
\end{flushleft}
\end{table}

By using a high-power lamp or laser to deliver intense and localized heating, rapid thermal annealing (RTA) can achieve temperature ramp rates as high as 400~\textdegree C/s \cite{Gunawan2004}. Typical applications include doping of silicon and repairing crystal lattices after ion implantation. It is also frequently used to improve material properties akin to conventional annealing. To date, there have only been two studies into annealing AlN waveguides, summarized in Table~\ref{tab:rta}. Dong \emph{et al}, have shown improvement in waveguide propagation loss at the mid-IR through conventional annealing in an oven, their optimal recipe was 400\textdegree C for 2~hr in an ambient gas environment \cite{Dong2019}. Wu \emph{et al.} have likewise found improvement by rapid thermal annealing of AlN waveguides when testing at 1550~nm. In their study, they found that repeated application of RTA would in many cases continuously improve loss up to a point. Their optimal recipe used a two part ramp, first reaching 500\textdegree C in 3~min with a 60~s soak then 800\textdegree C in an additional 3~min with a 90~s soak. The best loss was achieved by repeating this thermal cycle 8 times. In both studies they use a silicon substrate with \ce{SiO2} buried oxide and \ce{SiO2} cladding. To date this is the first study of rapid thermal annealing at shorter wavelengths, and the first with high quality AlN films on a sapphire substrate.

\begin{figure}[tb]
    \centering
    \includegraphics[width=\textwidth]{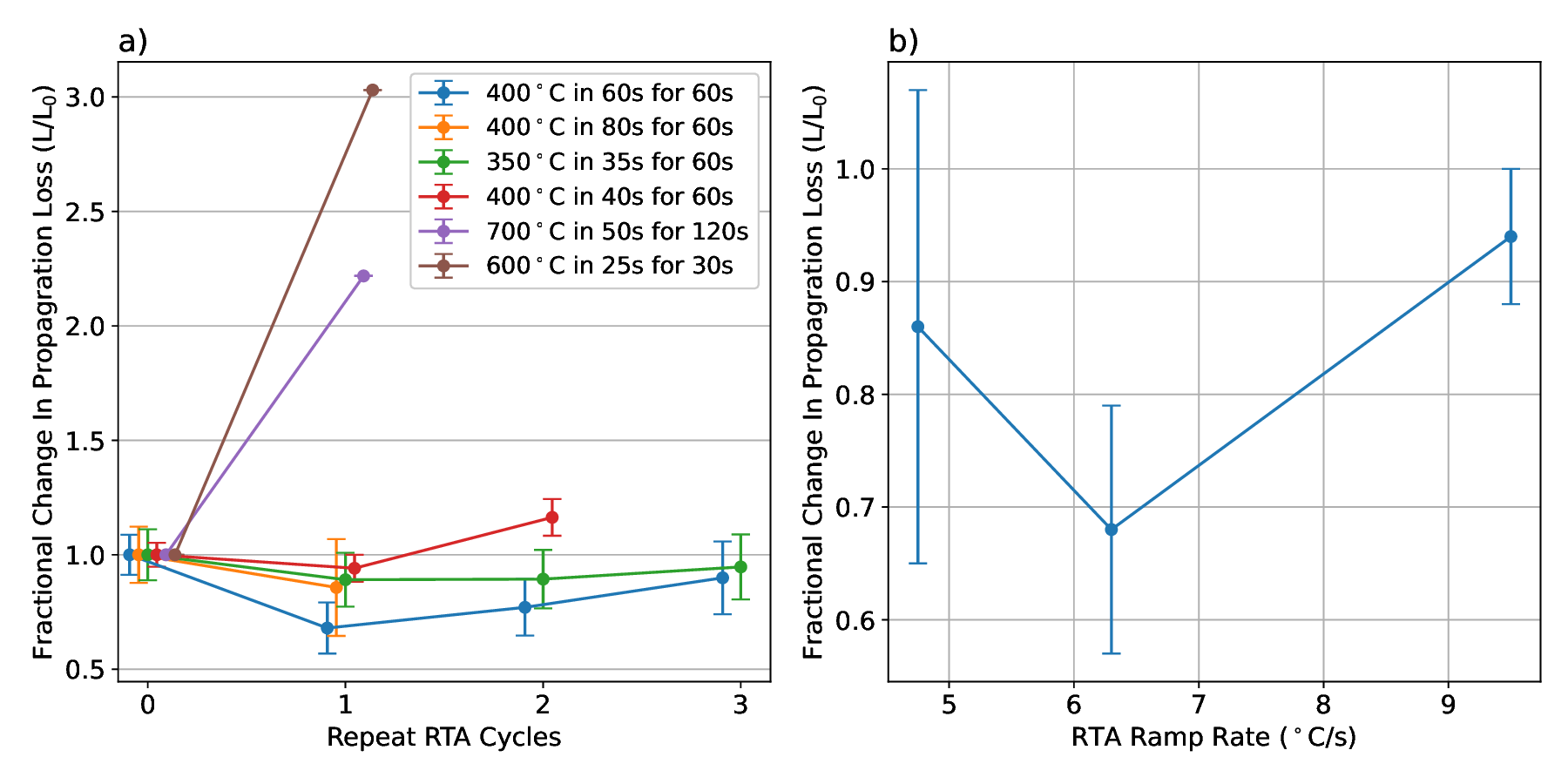}
    \caption{a) Fractional change in propagation loss, calculated as the ratio of the post-RTA loss over the pre-RTA loss. The same RTA recipe is applied until a worsening in the propagation loss is observed. b) Slicing of the data in a) to show the effect of ramp rate. Consistent across these data points is a ramp to 400~\textdegree C, and this fractional change is for the first time the RTA cycle is applied.}
    \label{fig:RTASummary}
\end{figure}

It is our finding that RTA, even for high quality films, can substantially improve the propagation loss. The normalized change in propagation loss for a range of maximum temperatures and ramp rates is shown in Fig.~\ref{fig:RTASummary}~a). The best improvement was found ramping to 400~\textdegree C which produced an average fractional improvement of 0.7$\pm$0.1, and a single chip best improvement of 0.45$\pm$0.07. This single chip improvement corresponds to the 2~dB/cm loss best performing device, demonstrating that even for ``good'' chips there is a benefit for RTA. This is half the optimal temperature found by Wu \emph{et al.} and more aligned with temperatures used for conventional annealing in an oven by Dong \emph{et al.} Consistently temperatures of 600~\textdegree C and higher led to significantly worse losses. Another divergence from Wu \emph{et al.} is the finding that multiple RTA cycles performed worse than a single cycle. Wu \emph{et al.} reached their maximum temperature at a maximum rate of 2.7~\textdegree C/s, while our slowest ramp is 4.7~\textdegree C/s, with the best improvement coming from a 6.3~\textdegree C/s, as shown in Fig.~\ref{fig:RTASummary}~b). This may explain the disparity between the two studies both in the optimal temperature and the efficacy of repeated cycles. At a maximum temperature of 350~\textdegree C the RTA appears to no longer affect the propagation loss even at a 9~\textdegree C /s ramp rate.

Not all of the samples tested were fabricated using the same recipe. Particularly relevant is the absence of an ALD passivation layer for the two earliest samples which coincide with the two worst fractional changes in propagation loss. In addition to the straightforward improvement in propagation loss from ALD passivation these results indicate there may be some benefit during RTA for the presence of \ce{Al2O3} passivation. All tested chips here are polycrystaline in the a-b axis and single crystal in the c-axis. All tests were done with samples which do not have electrodes, aluminum metal can diffuse into \ce{SiN_x} at temperatures as low as 450~\textdegree C \cite{Ogata1978}, and as a result RTA should be performed before electrodes are patterned.

These results demonstrate that even high-quality AlN waveguides on sapphire substrates benefit from rapid thermal annealing, with optimal improvement achieved by ramping to 400~\textdegree C in 60~s with a 60~s soak; this represents a relatively low temperature and fast ramp rate.

\section{Conclusion}\label{sec:conclusion}

Here, we have demonstrated a fabrication recipe which achieves propagation losses as low as 2.0$\pm$0.3~dB/cm at 852~nm. This recipe is on par with the best previously published loss numbers while also for the first time disclosing in detail all fabrication steps. Optimized electron beam lithography with 4x multi-pass, 4~nm shot pitch, and under-size over-dose shape PEC forms a foundation for the low loss waveguides. We show significant loss improvement through the incorporation of even a single ALD passivation layer, which has never been done with AlN waveguides to our knowledge. 10 ALD cycles are used for the best performing devices. We show further loss improvement by including a RTA step after waveguide patterning and cladding. This is the first ever investigation into RTA of high quality AlN on sapphire post cladding. We find that the optimal recipe, which can halve the propagation loss and on average produced 0.7$\pm$0.1 fractional improvement, is an RTA cycle to 400~\textdegree C with a 6.3~\textdegree C/s ramp and a 60~s soak. It is our intention that this paper facilitates improved fabrication recipes and improved losses in AlN waveguides to encourage more research into this promising material.

\section{Acknowledgments}
The University of Waterloo's QNFCF facility was used for this work. This infrastructure and the presented work would not be possible without the significant contributions of CFREF-TQT, CFI, ISED, the Ontario Ministry of Research \& Innovation and Mike \& Ophelia Lazaridis. Their support is gratefully acknowledged.

We further acknowledge Annabelle Wicentowich and Emma Rose Milne for their assistance with the literature review.

\newpage
\bibliographystyle{ieeetr}
\bibliography{sample}

\end{document}